\documentclass[11pt,twoside]{article}
\usepackage{asp2010}

\newcommand{\ha}{\ensuremath{\textrm{H} \alpha}}
\newcommand{\hii}{\ensuremath{\textrm{H II}}}

\newcommand{\kpc}{\ensuremath{\textrm{ kpc}}}
\newcommand{\pc}{\ensuremath{\textrm{ pc}}}

\newcommand{\s}{\ensuremath{\textrm{ s}}}

\newcommand{\K}{\ensuremath{\textrm{ K}}}

\newcommand{\arcdeg}{\ensuremath{^{\circ}}}

\newcommand{\EM}{\ensuremath{\textrm{EM}}}

\resetcounters

\begin{document}

\title{The role of turbulence in the warm ionized medium}
\author{Alex~S.~Hill$^1$, Kenneth~Wood$^{1,2}$, L.~Matthew Haffner$^1$}
\affil{$^1$Department of Astronomy, University of Wisconsin-Madison 53706 USA}
\affil{$^2$School of Physics \& Astronomy, University of St. Andrews, Scotland} 

\begin{abstract}
We discuss the role of turbulence in establishing the observed emission measures and ionization of the warm ionized medium. A Monte Carlo radiative transfer code applied to a snapshot of a simulation of a supernova-driven, multi-temperature, stratified ISM reproduces the essential observed features of the WIM.
\end{abstract}

In recent years, it has become clear that turbulence plays a central role in establishing the observed conditions in the warm ionized medium (WIM) of the Galaxy. Although the WIM was first detected four decades ago, the 1996 completion of the Wisconsin \ha\ Mapper (WHAM; \citealt{hrt03}) made possible large-scale studies of the physical conditions in the medium, particularly through the combination of \ha\ and collisionally excited metal lines. These and other studies have established that the WIM is nearly fully ionized \citep{rht98}, primarily by the O star Lyman continuum flux \citep{r90b}. However, the mechanism by which ionizing photons escape the immediate vicinity of the O stars near the Galactic plane to reach the observed $1.0-1.5 \kpc$ scale height of the WIM \citep{sw09} has been the subject of numerous investigations \citep{mc93,ds94,cbf02,wm04,hdb09}. These studies have generally found that the propagation of ionizing photons to $|z| \sim 1 \kpc$ in an ISM with the observed densities requires the presence of low-density paths through the ISM, but none produce the required low-density paths in a physically motivated way.

Meanwhile, the WIM has most often been modeled as consisting of either photoionized shells surrounding neutral clouds \citep{mo77} or as discrete clouds occupying some fraction of the total volume of the Galaxy \citep{r91}. However, this picture is clearly incomplete. The WIM has a high Reynolds number and therefore is turbulent \citep{b99}, as seen observationally \citep{ars95}. The recent advent of 3D simulations of supernova-driven turbulence in a multiphase ISM which cover dynamically significant time and size scales \citep{ab04,jm06} has made possible detailed studies of the interplay between the physical conditions revealed by optical emission lines and turbulence, helping to address the long-standing puzzle of the ionization of the medium.

\section{Distribution of emission measure}

\begin{figure}[bt]
\plotone{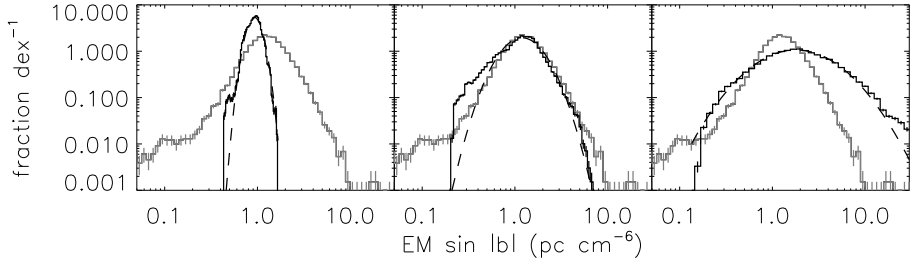}
\caption{{\em Grey lines:} Histogram of $\EM \sin |b|$ for diffuse WIM sightlines with $|b| > 10 \arcdeg$ from the WHAM survey, duplicated in all three panels. {\em Black lines:} Histogram of simulated \EM\ derived from $M_s \approx 0.7$ ({\em left}), $M_s \approx 2$ ({\em center}), and $M_s \approx 7$ ({\em right}) externally driven, isothermal MHD simulations. {\em Dashed lines:} lognormal fits to the simulated \EM\ distributions. From \citet{hbk08}.}
\label{fig:emhist}
\end{figure}

First, we consider the global distribution of \ha\ emission measures observed in the WHAM Northern Sky Survey \citep{hrt03}. We derived the emission measure, $\EM \equiv \int n_e^2 \,ds$, from the \ha\ surface brightness assuming a temperature of $8000 \K$. We then chose a subset of the WHAM survey which excludes sightlines towards large \hii\ regions associated with identified, discrete ionizing stars. We also excluded sightlines with Galactic latitude $|b| < 10 \arcdeg$, where extinction due to dust may affect the measured \ha\ surface brightness. The distribution of $\EM \sin |b|$ is well approximated by a lognormal distribution \citep[and Figure~\ref{fig:emhist}]{hbk08}.

\section{Simulations}

A random series of compressions and rarefactions, such as those that would occur in a turbulent fluid, establish a lognormal distribution of density \citep{v94}. This led us to consider 3D MHD simulations of artificially driven, isothermal turbulence \citep{klb07}. These simulations are generic and dimensionless. We applied physical scales appropriate to the WIM and derived simulated emission measure distributions \citep[and Figure~\ref{fig:emhist}]{hbk08}. In all cases, the simulated distributions are approximately lognormal. The width of the distribution is highly sensitive to the sonic Mach number ($M_s$) of the turbulence and is therefore the best parameter to compare to the observations. Subsonic models ($M_s \approx 0.7$) produce a narrower \EM\ distribution than observed, while highly supersonic ($M_s \approx 7$) models produce a wider \EM\ distribution than observed. However, the mildly supersonic ($M_s \approx 2$) simulations produce an \EM\ distribution consistent with observations.

\begin{figure}[bt]
\plotone{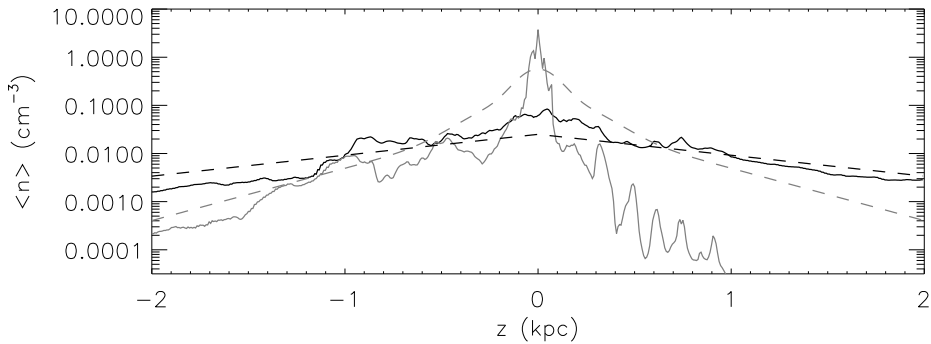}
\caption{Mean density of neutral ({\em solid grey line}) and ionized ({\em solid black line}) gas in the hydrodynamical-photoionization simulations as a function of height. \citeauthor{dl90} ({\em dashed grey line}) and exponential ({\em dashed black line}) density distributions for neutral and ionized gas are also shown. From \citet{whj10}.}
\label{fig:n_z}
\end{figure}

Although the isothermal, externally driven simulations reproduce the observed \EM\ distribution in the WIM, the driving mechanism is artificial. \citet{jm06} and \citet{jmb09} developed 3D hydrodynamical simulations of a $0.5 \kpc \times 0.5 \kpc \times 10 \kpc$ region of the ISM with parameterized heating and cooling and a Galactic gravitational potential. Supernova explosions stir the medium, driving turbulence and structure formation. These simulations produce a multi-phase, stratified ISM which is well suited for a more complete study of the role of turbulence in the WIM.

To investigate the WIM in the context of a supernova-driven turbulent ISM, we use a 3D Monte Carlo photoionization code \citep{wl00} to model the photoionization of gas in a snapshot of the Joung \& Mac Low supernova-driven models. We randomly place six O~stars in the simulated, initially neutral ISM with a Gaussian distribution and a scale height of $63 \pc$. The surface density of $24 \textrm{ O stars} \kpc^{-2}$ is chosen to represent a sample of O~stars near the sun \citep{gcc82}. We distribute the flux evenly between the O~stars, investigating total ionizing luminosities $Q$ over the range $10^{48} \s^{-1} \le Q \le 10^{50} \s^{-1}$. This span reflects a variation in efficiency of ionizing photons escaping from the immediate vicinity of the O~stars; the total ionizing flux available from six O~stars is  $\approx 10^{50} \s^{-1}$ \citep{vgs96}. While neutral hydrogen is opaque to Lyman continuum radiation, the optical depth of nearly fully ionized hydrogen is much lower. Therefore, we use an iterative process to establish a self-consistent distribution of neutral and ionized hydrogen.
Our results are presented by \citet{whj10} and summarized here. For ionizing luminosities $Q \ge 2 \times 10^{49} \s^{-1}$, ionizing photons reach the top of the grid at $|z| = 2 \kpc$. Figure~\ref{fig:n_z} shows the horizontally-averaged density of the neutral and ionized gas as a function of height for the $Q = 2 \times 10^{49} \s^{-1}$ case. At heights $|z| \gtrsim 1 \kpc$, the gas is nearly fully ionized.

\section{Conclusions}

Supernova-driven turbulent simulations combined with photoionization modeling produces the most self-consist modeling of the warm ionized medium to date. This finding is consistent with \citet{cbf02} and \citet{hdb09}, who found that analytic fractal density structures produce sufficient low-density pathways for the ionizing photons to travel large distances. In combination with energy budget arguments \citep{r90b}, these results strongly suggest that turbulence allows photoionization from hot stars to be the primary ionizing mechanism in the WIM, whereas a traditional picture of discrete clouds in the ISM does not allow the necessary propagation of ionizing photons at realistic ISM densities.

Inconsistencies between the combined hydrodynamical and photionization simulations and the observations suggest avenues for future investigation. The \citet{jmb09} hydrodynamical simulation produces a more centrally peaked gas distribution than observed, with densities higher than a \citet{dl90} distribution near the midplane and lower than Dickey-Lockman at $|z| \sim 0.2 \kpc$ \citep[Fig.~\ref{fig:n_z} and][]{whj10}, possibly partly due to effects of resolution or box size \citep{ab04}. The inclusion of magnetic fields in our supernova-driven simulations (work in progress) may help to address this inconsistency by applying additional vertical pressure, although \citet{ab05} found that the magnetic field in their simulations is largely uncorrelated with density.

Also, although our simulations reproduce the existence of the WIM at large heights without resorting to an {\em ad hoc} interstellar density structure, the distribution of emission measures is considerably wider than in the observations or the mildly supersonic artificially driven, isothermal simulations shown in Fig.~\ref{fig:emhist} and \citet{hbk08}. Therefore, the hydrodynamic supernova-driven simulations produce more contrast between low and high density gas than observed in the Galaxy. Because magnetic fields reduce the compressibility of the medium \citep{ab05}, we expect the inclusion of magnetic fields in the simulations to improve this inconsistency as well.

\acknowledgements The hydrodynamical simulations in this paper use the FLASH v2.4 code developed by the Center for Thermonuclear Flashes at the University of Chicago \citep{for00}. WHAM, A.~S.~H., and L.~M.~H. are supported by the NSF. We thank R.~Benjamin, R.~Joung, G.~Kowal, A.~Lazarian, and M.-M.~Mac Low for their contributions to the projects presented here.

\bibliography{hill_alex}

\end{document}